\documentclass[aps,pre,showpacs,superscriptaddress,floatfix, twocolumn]{revtex4}
\usepackage{graphicx}

\newcommand{\be}{\begin{equation}}
\newcommand{\ee}{\end{equation}}
\newcommand{\bea}{\begin{eqnarray}}
\newcommand{\eea}{\end{eqnarray}}

\begin{document}

\title{Constant flux relation for aggregation models with desorption and fragmentation}
\author{Colm Connaughton}
\email{connaughtonc@gmail.com} \affiliation {Center for Nonlinear Studies,
Los Alamos National Laboratory, Los Alamos, NM 87545, USA}
\author{R. Rajesh}
\email{rrajesh@imsc.res.in}
\affiliation{
Institute of Mathematical Sciences, CIT Campus, Taramani, Chennai-600013,
India}
\author{Oleg Zaboronski}
\email{olegz@maths.warwick.ac.uk}
\affiliation{Mathematics Institute, University of Warwick, Gibbet Hill
Road, Coventry CV4 7AL, UK}

\date{\today}

\begin{abstract}
We study mass fluxes in aggregation models where mass transfer to
large scales by aggregation occurs alongside desorption or fragmentation.
Two models are considered. (1) A system of diffusing, aggregating
particles with influx and outflux of particles (in-out model) (2) A system of 
diffusing aggregating particles with fragmentation (chipping model). Both these
models can exist in phases where probability distributions are power laws. In
these power law phases, we argue that
the two point correlation function should have a 
certain homogeneity exponent. These arguments are based on the
exact constant flux scaling valid for simple aggregation with input.
Predictions are compared with  Monte Carlo simulations.
\end{abstract}
\pacs{05.20.-y, 47.27.-i, 47.35.Bb, 61.43.Hv}

\maketitle

\section{\label{introduction} Introduction}

A variety of aggregation--diffusion models have been constructed over the
years by defining simple stochastic rules governing the evolution of a 
set of particles on a discrete lattice. Some of the interest in models of 
this type comes from the fact that they can be considered as minimal models
of physical aggregation diffusion systems which provide a theoretical 
framework in which to study such phenomena as flocculation and gelation in
aerosols and emulsions \cite{DRA1972}. In addition, certain aggregation models have been 
shown to be related to models describing seemingly unrelated things such as 
the geometry of river networks \cite{huber1991}, the distribution of
forces in granular media \cite{coppersmith1996} or the directed abelian 
sandpile model \cite{liu1995}. In analysing
such models, most of the theoretical effort has focussed on determining
the average mass density, $P(m,t)$. In situations where a mean-field 
description is appropriate, this can be calculated by solving a 
Smoluchowski--like kinetic equation \cite{LEY2003}. However, many of the applications
are in low spatial dimensions where mean-field theory is
typically inapplicable \cite{KAN1984,privman1997,KRZ2002}. 
In such situations, diffusive fluctuations dominate,
rendering the determination of $P(m,t)$ quite 
non-trivial. Nevertheless, much progress has been  made for specific models.

Despite the fact that fluctuations lead to non-trivial
statistics for the mass distribution, very little is known, even 
numerically, about higher order correlation functions
\cite{munasinghe2006a,munasinghe2006b}. This  
is the issue which we would like to address here. We consider a subclass of
aggregation--diffusion models, specifically those which include deposition
of monomers. Such models reach a statistically stationary state where the 
input of small masses is balanced by the depletion of small masses to generate
larger ones by aggregation. In recent work \cite{CRZtmlong,CRZcfr} we argued that it is useful to
think of this stationary state as analogous to the stationary state of a 
turbulent system with the mass flux playing the role of the energy flux. One
of the most important fruits of this analogy is the realisation that 
aggregation models having a stationary state with constant mass flux must
satisfy an analogue of Kolmogorov's 4/5 Law \cite{KOL1941c,frischbook}. This constraint, which we
call the Constant Flux Relation (CFR), fixes exactly the scaling of a 
special correlation function of the mass distribution, namely the one
which carries the mass flux. The power of this result is that it
does not require any mean-field assumptions and holds equally well in the
fluctuation dominated regime. In fact, it determines exactly the scaling
of the flux-carrying correlation function for a broad class of homogeneous
aggregation kernels even if the scaling of $P(m,t)$ itself is not known. 
While the determination of a single correlation function is a modest step
when faced with the problem of determining the full statistics, it can 
nevertheless be a powerful marker. For example, knowledge of the CFR exponent
allowed us to give a relatively simple proof of the multifractality of the
mass distribution for constant kernel aggregation in one dimension \cite{CRZtmshort}.

In our earlier paper \cite{CRZcfr}, we derived CFR for a simple aggregation--diffusion 
model which we referred to as the Mass Model (MM). It was a model
of diffusing particles undergoing mass conserving aggregation along with
input of particles of small mass. The cascade was in the mass space with
driving at small mass scales, dissipation at infinity, conserved quantity 
being mass and aggregation the process transferring mass to larger mass scales.
We also derived analogous results for a collection of other models, aiming
to stress the ubiquity of the approach and the central role played by
conservation laws. In this paper we restrict ourselves to the implications of
CFR for aggregation models. For clarity we will restate the argument for the 
MM.  In addition, we ask whether the argument leading to CFR for the MM can
tell us anything when the conservation law is less obvious, or when the  
driving and dissipation scales are not widely separated as in the MM.  We
address this question in the context of two models studied earlier in relation
to  nonequilibrium phase transitions \cite{MKB1}.
These  models, which we call the in-out model and the chipping model, are 
generalisations of the MM. In both, the simple transfer
of mass in the MM is disrupted all mass scales by evaporation in
the former model and by fragmentation in the latter. In addition, the source of
small mass scales is generated from within for the chipping model and not 
controlled from outside.  The analysis for the MM was exact. Here we proceed 
by analogy in cases where an exact approach is yet to be developed, backing 
up our heuristic arguments with numerical simulations.

The rest of the paper is organised as follows. In Sec.~\ref{model} the 
two models are defined and known results are briefly reviewed.
Sec.~\ref{sec_cfr} describes the CFR for these models, and predicts the
scaling behaviour of the two point joint probability distribution function. 
Sec.~\ref{numerics} contains the results of Monte Carlo simulations.
Sec.~\ref{summary} contains a summary and conclusions.

\section{\label{model} Models and Review}

In this section, the 
two models- in-out model and chipping model - are defined and relevant
earlier results are reviewed.
The in-out model describes a system
of diffusing, aggregating particles with influx and outflux of particles. 
The chipping model describes a closed system
of diffusing, aggregating particles with fragmentation.

We define the models on a one dimensional lattice with
periodic boundary conditions; generalisations to $d$-dimensional
hyper-cubic lattice is straight forward.

\subsection{In-out model}

Each site $i$ of the lattice has a non-negative
integer mass variable $m_i \geq 0$.  Given a certain configuration of
masses at time $t$, the system evolves in an infinitesimal time $dt$ as
follows. A site $i$ is chosen at random (with probability $dt$), and then
the following events
can occur.  (i) Adsorption: with probability $q/(p+q+1)$, unit mass is
adsorbed at site $i$; thus $m_i \rightarrow m_i+1$.  (ii) Desorption: if
the mass $m_i$ is greater than zero, then with probability $p/(p+q+1)$,
unit mass is desorbed from site $i$; thus $m_i \rightarrow m_i -1$
provided $m_i \geq 0$.  (iii) Diffusion and aggregation: 
with probability $1/(p+q+1)$
the mass $m_i$ moves to a randomly chosen nearest neighbour;
thus $m_i \rightarrow 0$ and $m_{i\pm 1} \rightarrow m_{i\pm 1}
+ m_i$.  The initial condition is chosen to be to be one in $m_i=0$ for all $i$. 
The model has two parameters, $p,q$. We shall refer
to this model as the in-out model.

Let $P(m,t)$ denote probability that a site has mass $m$ at time $t$. The
large time limit will be denoted by $P(m)$, i.e., $P(m) = \lim_{t\rightarrow \infty} P(m,t)$. 
When the adsorption rate $q$ is increased keeping the desorption 
rate $p$ fixed,
the system undergoes a nonequilibrium phase transition across a critical
line $q_c(p)$ from a phase in which $P(m)$
has an exponential tail to one in which it has an algebraic tail
for large mass; i.e,
\be
P(m) \sim \cases{
e^{-m/m^*} & when $q < q_c(p)$, \cr
m^{-\tau_c} & when $q = q_c(p)$, \cr
m^{-\tau} & when $q > q_c(p)$, \cr}
\label{eq:1}
\ee
where $m^*$ is a $q$ dependent cut-off, and $\tau$ and $\tau_c$ are
exponents characterising the power law decay \cite{MKB1,MKB2000}. 
The three phases will be called
as the exponential phase ($q<q_c$), the critical phase
($q=q_c$) and the growing phase ($q>q_c$).
In addition, it was argued
that as a function of the small deviation $\tilde{q} = q-q_c$, and large
time $t$, $P(m, \tilde{q}, t)$ displays the scaling form
\be
P(m,\tilde{q}, t) \sim \frac{1}{m^{\tau_c}} Y \left( m \tilde{q}^\phi,
\frac{m}{t^\alpha} \right),
\label{eq:2}
\ee
in terms of three unknown exponents $\phi$, $\alpha$, $\tau_c$, and
the
two variable scaling function $Y$. The three exponents were determined in
all dimensions \cite{rajesh2004}. Of interest in this paper is the 
behaviour in one dimension wherein \cite{rajesh2004}
\bea
P(m,t) &\sim& \frac{1}{m^{11/6}} f_c\left(\frac{m}{t^{3/5}} \right), \quad
q=q_c(p), \\
P(m,t) &\sim& \frac{1}{m^{4/3}} f_g\left(\frac{m}{t^{3/2}} \right), \quad
q\gg q_c(p), 
\eea
where the scaling functions $f_c(x), f_g(x) \rightarrow x^0$ when
$x\rightarrow 0$ and $f_c(x), f_g(x) \rightarrow 0$ when $x\rightarrow
\infty$.

\subsection{Chipping model}

Each site $i$ of the lattice has a non-negative
integer mass variable $m_i \geq 0$.  Given a certain configuration of
masses at time $t$, 
the system evolves in
infinitesimal time $dt$ as follows.
A site $i$ is chosen at random (with probability $dt$), and then
the following events can occur.
(i) Chipping: with probability $w/(w+1)$, unit mass is
chipped out from site $i$ and added to a neighbour; thus $m_i \rightarrow m_i-1$
and $m_{i\pm 1} \rightarrow m_{i\pm 1}
+ 1$.  (ii) Diffusion and aggregation: with probability $1/(1+w)$, the mass $m_i$
moves to a randomly chosen nearest neighbour; 
thus $m_i \rightarrow 0$ and $m_{i\pm 1} \rightarrow m_{i\pm 1}
+ m_i$. The initial condition is chosen to be to be one in which density
is uniform.
The model has two parameters: $\rho$, the mean density and
$w$,  the ratio of the chipping rate to the hopping rate.
We shall refer
to this model as the chipping model.

The system undergoes a transition
in the $\rho$-$w$ plane\cite{MKB1,MKB2}. There is a critical 
line $\rho_c(w)$ in
the $\rho$-$w$ plane that separates two types of asymptotic behaviours of
$P(m)$.  For fixed $w$, as $\rho$ is varied across the critical value
$\rho_c(w)$, the large $m$ behaviour of $P(m)$ was found to be,
\be
P(m)\sim \cases
{e^{-m/m^{*}} &$\rho<\rho_c(w)$,\cr
m^{-\tau} &$\rho=\rho_c(w)$,\cr
m^{-\tau}+ {\rm infinite\,\,\, aggregate} &$\rho>\rho_c(w)$.\cr}
\label{pm}
\ee

Thus, the tail of the mass distribution changes from having an
exponential decay to an algebraic decay as $\rho$ approaches $\rho_c$ from
below. As one increases $\rho$ beyond $\rho_c$, this asymptotic algebraic
part of the critical distribution remains unchanged but in addition an
infinite aggregate forms. This means that all the additional mass
$(\rho-\rho_c)V$ (where $V$ is the volume of the system) condenses onto a
single site and does not disturb the background critical distribution.
This is analogous, in spirit, to the condensation of a macroscopic number
of bosons onto the single $k=0$ mode in an ideal Bose gas as the
temperature goes below a certain critical value. 

The critical density $\rho_c(w)$ was found to be \cite{rajeshmkb1}
\be
\rho_c(w)=\sqrt{w+1}-1,
\ee
in all dimensions. In addition, it was argued that the exponent $\tau$ is
super-universal and was equal to $5/2$ in all dimensions. In particular,
it was argued that in one dimension
\be
P(m,t) \sim \frac{1}{m^{5/2}} f_{ch}\left( \frac{m}{t^{1/3}} \right),
\ee
where the scaling functions $f_{ch}(x) \rightarrow x^0$ when
$x\rightarrow 0$ and $f_{ch}(x) \rightarrow 0$ when $x\rightarrow
\infty$.

\section{\label{sec_cfr} Constant Flux Relation}

In this section, we summarise the CFR argument specifically for aggregation 
models. We then apply it to the in-out model and the chipping model in a  
heuristic way.  The heuristic steps to CFR are as follows.
First identify the conserved quantity and the space in which it flows.
Second, use the equation of motion to write a Boltzmann like 
continuity equation for the average density of the conserved quantity in this 
space.  This equation identifies a flux-carrying correlation function, $\Pi$, 
and a nonlinear coupling, $T$, controlling the flow among degrees of freedom.
Dimensional analysis may be used to determine the scaling of $\Pi$ 
corresponding to constant flux. In Ref.\cite{CRZcfr}, we showed how to make
this dimensional argument exact. In this paper we will restrict ourselves
to aggregation problems where the conserved quantity is mass, or potentially
a power of mass. contenting ourselves with a dimensional derivation.

Using the methods of Refs.\cite{zaboronski2001, CRZtmlong}, the 
mass density in the MM is controlled by an equation with the following 
structure:
\begin{equation}
\label{simpleSSE}
\partial_t(m P(m,t)) = \int T_{m,m_1,m_2} \Pi_{m,m_1,m_2} d m_1 d m_2 +\ldots
\end{equation}
where $T_{m,m_1,m_2}$ is the aggregation kernel and $\Pi_{m,m_1,m_2}$ is
the flux-carrying correlation function of interest here. In 
Refs.\cite{CRZtmlong,CRZtmshort} we showed  that $\Pi_{m,m_1,m_2}$ takes the 
form $ \Pi_{m,m_1,m_2} = m P(m_1,m_2,t) \delta(m-m_1-m_2)$ where the quantity 
$P(m_1,m_2,t)$ should be interpreted physically as the probability of finding
particles of masses $m_1$ and $m_2$ on adjacent lattice sites.
Of course, when written out properly, Eq.(\ref{simpleSSE}) is much more
complex, with  the ``$\ldots$'' representing additional integrals of the
form shown (but with permuted arguments), a diffusion term, a 
source term and a noise term. The diffusion term is neglected since it is
zero on average for a statistically homogeneous system. Likewise the noise,
having mean zero, is neglected on average. The source term is zero for
large masses.  We can then take Eq.(\ref{simpleSSE}) to define the flux
of mass, $J(m,t)$, in mass space: 
\begin{equation}
\label{simpleJm}
\partial_m J(m,t) \equiv \int T_{m,m_1,m_2} \Pi_{m,m_1,m_2} d m_1 d m_2.
\end{equation}
Given that the aggregation kernel is homogeneous of degree, $\beta$,
and assuming that the flux-carrying correlation function is homogeneous of
degree $h$, it follows that the flux itself scales as $J \sim m^{\beta+h+3}$.
The essential observation is that a stationary state of Eq.(\ref{simpleSSE})
corresponds, from Eq.(\ref{simpleJm}), to a constant flux of mass, i.e. $J(m)$
independent of $m$. This immediately gives the CFR scaling for the 
flux-carrying correlation function, $\Pi_{m,m_1,m_2}$:
\begin{equation}
\label{CFRExponentMM}
h=-\beta-3.
\end{equation}
Hence, in the stationary state, we expect that, for large masses,
$m P(m_1,m_2) \delta(m-m_1-m_2) \sim m^{-\beta-3}$
or,
\begin{equation}
\label{cfr}
P(m_1,m_2)\sim m^{-\beta-3}.
\end{equation}
Suppose, instead of mass, the conserved quantity was a power of the mass, 
$m^\gamma$. Formally, this argument would lead us to expect that, in 
the stationary 
state,
$m^\gamma P(m_1,m_2) \delta(m-m_1-m_2) \sim m^{-\beta-3}$.
This  would give
\begin{equation}
\label{pseudocfr}
P(m_1,m_2)\sim m^{-\gamma-\beta-2},
\end{equation}
which will be relevant to the subsequent discussion.

In the MM, and the other cases considered in Ref.\cite{CRZcfr}, the conserved quantity
was obvious. In the in-out model and the chipping model, it is not clear 
what is conserved, and whether there is a cascade in some quantity. 
In both cases, if we look in the mass space, then there is 
loss of mass at all mass scales
either through desorption or through chipping. In addition, the flux in the chipping model is
created from within through the fragmentation process. 
To apply Eq.~(\ref{cfr}) or  Eq.~(\ref{pseudocfr})to the two models being considered, we do the following. Identify $I$
by using the known behaviour of $P(m,t)$. Let $P(m,t)$ have the scaling form
\be
P(m,t) = \frac{1}{m^{\tau}} F\left(\frac{m}{t^{\delta}} \right)
\ee 
where the scaling function $F(x)$ goes to a constant for small $x$ and to zero for large $x$.
We identify $I$ to be $I=m^{\gamma}$, where $\gamma$ is fixed by the condition that $\langle I \rangle \sim t$. Thus,
\be
\gamma = \frac{1}{\delta} +\tau-1.
\ee
%Now, in both the models considered the relevant correlation function is the two-point distribution
%$P(m_1,m_2)$, the probability of having masses $m_1$ and $m_2$ on neighbouring sites. CFR then says that
%\be
%m^{\gamma} P(m,m) \delta(m) \sim m^{-3-\beta}
%\ee
%or
%\be
%P(m_1,m_2) \sim \frac{1}{(m_1 m_2)^{(2+\gamma+\beta)/2}}
%\ee

For both the models considered $\beta=0$. In one dimension, for the in-out
model $\gamma = 5/2$  at $q=q_c(p)$ and $\gamma=1$ for $q\gg q_c(p)$. For the
chipping model in one dimension $\gamma=9/2$ at $\rho=\rho_c(w)$.
Thus,
\bea
P(m_1,m_2) &\sim & (m_1 m_2)^{-9/4},  q=q_c(p),
\mbox{in-out}, d=1  \label{cfrinout}\\
P(m_1,m_2) &\sim & (m_1 m_2)^{3}, ~ q\gg q_c(p),
~ \mbox{in-out}, ~d=1  \label{cfrinoutgrow}\\
P(m_1,m_2) &\sim & (m_1 m_2)^{-13/4}, \rho=\rho_c, \mbox{chipping}, d=1. 
\label{cfrchipping}
\eea

\section{\label{numerics} Monte Carlo Simulations}

In this section, we present simulation results for the in-out and chipping
models and compare the results with the predictions in
Eqs.~(\ref{cfrinout}) and (\ref{cfrchipping}).  
We first present results for the in-out model. CFR predicts that the joint
distribution function $P(m_i,m_{i+1}) \sim (m_i m_{i+1})^{-9/4}$. In
simulations what we will measure is the quantity
\be
\pi(m) = \int_m^{\infty} dm' P(m,m').
\ee
CFR then predicts that 
\bea
\pi(m) & \sim & m^{-7/2}, q=q_c(p)\quad \mbox{in-out model},~d=1,\\
\pi(m) & \sim & m^{-2}, q\gg q_c(p)\quad \mbox{in-out model},~d=1.
\eea

\begin{figure}
\includegraphics[width=\columnwidth]{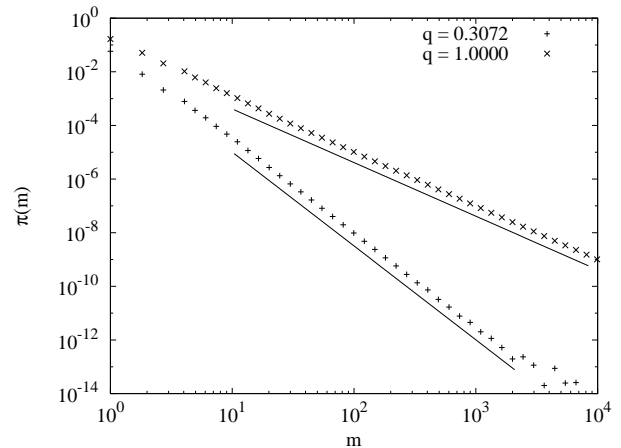}
\caption{\label{fig1} The variation of $\pi(m)$ with
$m$ is shown for different $q$'s in the in-out model.
The lower straight line has an exponent $-3.5$ and
corresponds to $q=q_c(p)$. The upper
straight line has an exponent $-2.0$ and corresponds to
the growing phase $q \gg q_c(p)$.
}
\end{figure}

We did Monte Carlo simulations on a one dimensional lattice of size $L=4096$.
The desorption rate $p$ is fixed to be one. For $p=1$, $q_c\approx 0.3072$.
In Fig.~\ref{fig1}, we show the variation of $\pi(m)$ with $m$ for
$q=0.3072$.
As can be seen, the exponent $-3.5$ is a good fit, thus
consistent with CFR. Also shown in Fig.~\ref{fig1} is $\Pi(m)$ for $q=1.0$.
For this value of $q$, the system is in the growing phase, and CFR predicts
that $\Pi(m) \sim m^{-2}$. Again, this is borne out by simulations (see top
curve of Fig.~\ref{fig1}).

\begin{figure}
\includegraphics[width=\columnwidth]{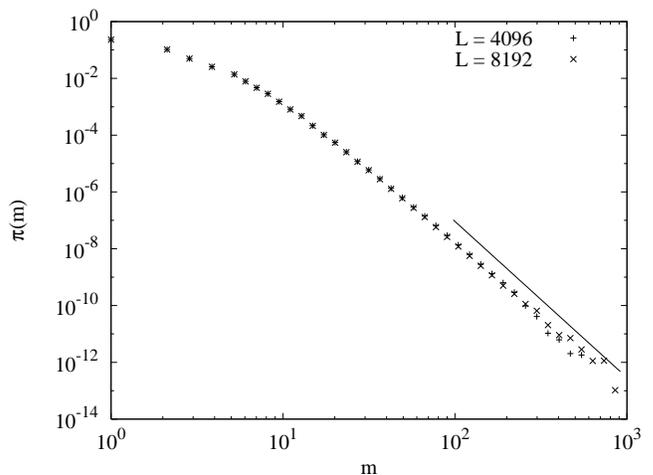}
\caption{\label{fig2} The variation of $\pi(m)$ with
$m$ is shown for two different lattice sizes in the chipping model.
$\rho=\rho_c$.
The straight line has an exponent $-5.5$.
}
\end{figure}

For the chipping model, CFR predicts that at $\rho=\rho_c(w)$,
\be
\pi(m) \sim m^{-11/2}, \quad \mbox{chipping model},~d=1.
\ee
Monte Carlo simulations were done for lattices of size $4096$ and $8192$.
The chipping rate $w=24.0$. $\rho$ was chosen to be $\rho=\rho_c=4.0$.
The results are shown in Fig.~\ref{fig2}. The data is consistent with the CFR
predictions.

\section{\label{summary} Summary and Conclusions}

In this paper, we examined the consequences of a constant flux on two 
models where there were no obvious conserved quantities and the 
dissipation and driving scales were not widely separated.  The two models
considered were examples of systems undergoing nonequilibrium phase transitions.
CFR holds in the phases where the probability distributions are power laws. The CFR
prediction was borne out by numerical simulations. However, an analytic approach is lacking.
For this purpose, one possibly needs to work with the effective field theories for
these models.

For the chipping model, the fact that the two point correlations have a different exponents
from the mean field answer clearly shows that the super-universality of $P(m)$ is a 
coincidence. It is an open problem as to why this exponent does not change with
dimension.

%\bibliography{ref}

\begin{thebibliography}{21}
\expandafter\ifx\csname natexlab\endcsname\relax\def\natexlab#1{#1}\fi
\expandafter\ifx\csname bibnamefont\endcsname\relax
  \def\bibnamefont#1{#1}\fi
\expandafter\ifx\csname bibfnamefont\endcsname\relax
  \def\bibfnamefont#1{#1}\fi
\expandafter\ifx\csname citenamefont\endcsname\relax
  \def\citenamefont#1{#1}\fi
\expandafter\ifx\csname url\endcsname\relax
  \def\url#1{\texttt{#1}}\fi
\expandafter\ifx\csname urlprefix\endcsname\relax\def\urlprefix{URL }\fi
\providecommand{\bibinfo}[2]{#2}
\providecommand{\eprint}[2][]{\url{#2}}

\bibitem[{\citenamefont{Drake}(1972)}]{DRA1972}
\bibinfo{author}{\bibfnamefont{R.}~\bibnamefont{Drake}}, in
  \emph{\bibinfo{booktitle}{Topics in Current Aerosol Research (part 2), Vol.
  3}}, edited by \bibinfo{editor}{\bibfnamefont{G.}~\bibnamefont{Hidy}}
  \bibnamefont{and} \bibinfo{editor}{\bibfnamefont{R.}~\bibnamefont{Brock}}
  (\bibinfo{publisher}{Pergamon Press}, \bibinfo{address}{New York},
  \bibinfo{year}{1972}), p. \bibinfo{pages}{201}.

\bibitem[{\citenamefont{Huber}(1991)}]{huber1991}
\bibinfo{author}{\bibfnamefont{G.}~\bibnamefont{Huber}},
  \bibinfo{journal}{Physica A} \textbf{\bibinfo{volume}{170}},
  \bibinfo{pages}{463} (\bibinfo{year}{1991}).

\bibitem[{\citenamefont{Coppersmith et~al.}(1996)\citenamefont{Coppersmith,
  h~Liu, Majumdar, Narayan, and Witten}}]{coppersmith1996}
\bibinfo{author}{\bibfnamefont{S.~N.} \bibnamefont{Coppersmith}},
  \bibinfo{author}{\bibfnamefont{C.}~\bibnamefont{h~Liu}},
  \bibinfo{author}{\bibfnamefont{S.}~\bibnamefont{Majumdar}},
  \bibinfo{author}{\bibfnamefont{O.}~\bibnamefont{Narayan}}, \bibnamefont{and}
  \bibinfo{author}{\bibfnamefont{T.~A.} \bibnamefont{Witten}},
  \bibinfo{journal}{Phys. Rev. E} \textbf{\bibinfo{volume}{53}},
  \bibinfo{pages}{4673} (\bibinfo{year}{1996}).

\bibitem[{\citenamefont{h~Liu et~al.}(1995)\citenamefont{h~Liu, Nagel,
  Schecter, Coppersmith, Majumdar, Narayan, and Witten}}]{liu1995}
\bibinfo{author}{\bibfnamefont{C.}~\bibnamefont{h~Liu}},
  \bibinfo{author}{\bibfnamefont{S.}~\bibnamefont{Nagel}},
  \bibinfo{author}{\bibfnamefont{D.~A.} \bibnamefont{Schecter}},
  \bibinfo{author}{\bibfnamefont{S.~N.} \bibnamefont{Coppersmith}},
  \bibinfo{author}{\bibfnamefont{S.~N.} \bibnamefont{Majumdar}},
  \bibinfo{author}{\bibfnamefont{O.}~\bibnamefont{Narayan}}, \bibnamefont{and}
  \bibinfo{author}{\bibfnamefont{T.~A.} \bibnamefont{Witten}},
  \bibinfo{journal}{Science} \textbf{\bibinfo{volume}{269}},
  \bibinfo{pages}{513} (\bibinfo{year}{1995}).

\bibitem[{\citenamefont{Leyvraz}(2003)}]{LEY2003}
\bibinfo{author}{\bibfnamefont{F.}~\bibnamefont{Leyvraz}},
  \bibinfo{journal}{Phys. Rep.} \textbf{\bibinfo{volume}{383}},
  \bibinfo{pages}{95} (\bibinfo{year}{2003}).

\bibitem[{\citenamefont{Kang and Redner}(1984)}]{KAN1984}
\bibinfo{author}{\bibfnamefont{K.}~\bibnamefont{Kang}} \bibnamefont{and}
  \bibinfo{author}{\bibfnamefont{S.}~\bibnamefont{Redner}},
  \bibinfo{journal}{Phys. Rev. A} \textbf{\bibinfo{volume}{30}},
  \bibinfo{pages}{2833} (\bibinfo{year}{1984}).

\bibitem[{\citenamefont{Privman}(1997)}]{privman1997}
\bibinfo{editor}{\bibfnamefont{V.}~\bibnamefont{Privman}}, ed.,
  \emph{\bibinfo{title}{Nonequilibrium Statistical Mechanics in one
  dimensions}} (\bibinfo{publisher}{Cambridge University Press},
  \bibinfo{address}{Cambridge}, \bibinfo{year}{1997}).

\bibitem[{\citenamefont{Krishnamurthy et~al.}(2002)\citenamefont{Krishnamurthy,
  Rajesh, and Zaboronski}}]{KRZ2002}
\bibinfo{author}{\bibfnamefont{S.}~\bibnamefont{Krishnamurthy}},
  \bibinfo{author}{\bibfnamefont{R.}~\bibnamefont{Rajesh}}, \bibnamefont{and}
  \bibinfo{author}{\bibfnamefont{O.}~\bibnamefont{Zaboronski}},
  \bibinfo{journal}{Phys. Rev. E} \textbf{\bibinfo{volume}{66}},
  \bibinfo{pages}{066118} (\bibinfo{year}{2002}).

\bibitem[{\citenamefont{Munasinghe
  et~al.}(2006{\natexlab{a}})\citenamefont{Munasinghe, Rajesh, and
  Zaboronski}}]{munasinghe2006a}
\bibinfo{author}{\bibfnamefont{R.}~\bibnamefont{Munasinghe}},
  \bibinfo{author}{\bibfnamefont{R.}~\bibnamefont{Rajesh}}, \bibnamefont{and}
  \bibinfo{author}{\bibfnamefont{O.}~\bibnamefont{Zaboronski}},
  \bibinfo{journal}{Phys. Rev. E} \textbf{\bibinfo{volume}{73}},
  \bibinfo{pages}{051103} (\bibinfo{year}{2006}{\natexlab{a}}).

\bibitem[{\citenamefont{Munasinghe
  et~al.}(2006{\natexlab{b}})\citenamefont{Munasinghe, Rajesh, Tribe, and
  Zaboronski}}]{munasinghe2006b}
\bibinfo{author}{\bibfnamefont{R.}~\bibnamefont{Munasinghe}},
  \bibinfo{author}{\bibfnamefont{R.}~\bibnamefont{Rajesh}},
  \bibinfo{author}{\bibfnamefont{R.}~\bibnamefont{Tribe}}, \bibnamefont{and}
  \bibinfo{author}{\bibfnamefont{O.}~\bibnamefont{Zaboronski}},
  \bibinfo{journal}{Commun. Math. Phys} \textbf{\bibinfo{volume}{268}},
  \bibinfo{pages}{717} (\bibinfo{year}{2006}{\natexlab{b}}).

\bibitem[{\citenamefont{Connaughton et~al.}(2006)\citenamefont{Connaughton,
  Rajesh, and Zaboronski}}]{CRZtmlong}
\bibinfo{author}{\bibfnamefont{C.}~\bibnamefont{Connaughton}},
  \bibinfo{author}{\bibfnamefont{R.}~\bibnamefont{Rajesh}}, \bibnamefont{and}
  \bibinfo{author}{\bibfnamefont{O.}~\bibnamefont{Zaboronski}},
  \bibinfo{journal}{Physica D} \textbf{\bibinfo{volume}{222}},
  \bibinfo{pages}{97} (\bibinfo{year}{2006}).

\bibitem[{\citenamefont{Connaughton et~al.}(2007)\citenamefont{Connaughton,
  Rajesh, and Zaboronski}}]{CRZcfr}
\bibinfo{author}{\bibfnamefont{C.}~\bibnamefont{Connaughton}},
  \bibinfo{author}{\bibfnamefont{R.}~\bibnamefont{Rajesh}}, \bibnamefont{and}
  \bibinfo{author}{\bibfnamefont{O.}~\bibnamefont{Zaboronski}},
  \bibinfo{journal}{Phys. Rev. Lett} \textbf{\bibinfo{volume}{98}},
  \bibinfo{pages}{080601} (\bibinfo{year}{2007}).

\bibitem[{\citenamefont{Kolmogorov}(1941)}]{KOL1941c}
\bibinfo{author}{\bibfnamefont{A.~N.} \bibnamefont{Kolmogorov}},
  \bibinfo{journal}{Dokl. Akad. Nauk SSSR} \textbf{\bibinfo{volume}{32}},
  \bibinfo{pages}{15} (\bibinfo{year}{1941}), \bibinfo{note}{reprinted in Proc.
  Roy. Soc. Lond. A, 434, 15-17, (1991)}.

\bibitem[{\citenamefont{Frisch}(1995)}]{frischbook}
\bibinfo{author}{\bibfnamefont{U.}~\bibnamefont{Frisch}},
  \emph{\bibinfo{title}{Turbulence: The Legacy of A. N. Kolmogorov}}
  (\bibinfo{publisher}{Cambridge University Press},
  \bibinfo{address}{Cambridge}, \bibinfo{year}{1995}).

\bibitem[{\citenamefont{Connaughton et~al.}(2005)\citenamefont{Connaughton,
  Rajesh, and Zaboronski}}]{CRZtmshort}
\bibinfo{author}{\bibfnamefont{C.}~\bibnamefont{Connaughton}},
  \bibinfo{author}{\bibfnamefont{R.}~\bibnamefont{Rajesh}}, \bibnamefont{and}
  \bibinfo{author}{\bibfnamefont{O.}~\bibnamefont{Zaboronski}},
  \bibinfo{journal}{Phys. Rev. Lett} \textbf{\bibinfo{volume}{94}},
  \bibinfo{pages}{194503} (\bibinfo{year}{2005}).

\bibitem[{\citenamefont{Majumdar et~al.}(1998)\citenamefont{Majumdar,
  Krishnamurthy, and Barma}}]{MKB1}
\bibinfo{author}{\bibfnamefont{S.~N.} \bibnamefont{Majumdar}},
  \bibinfo{author}{\bibfnamefont{S.}~\bibnamefont{Krishnamurthy}},
  \bibnamefont{and} \bibinfo{author}{\bibfnamefont{M.}~\bibnamefont{Barma}},
  \bibinfo{journal}{Phys. Rev. Lett} \textbf{\bibinfo{volume}{81}},
  \bibinfo{pages}{3691} (\bibinfo{year}{1998}).

\bibitem[{\citenamefont{Majumdar
  et~al.}(2000{\natexlab{a}})\citenamefont{Majumdar, Krishnamurthy, and
  Barma}}]{MKB2000}
\bibinfo{author}{\bibfnamefont{S.~N.} \bibnamefont{Majumdar}},
  \bibinfo{author}{\bibfnamefont{S.}~\bibnamefont{Krishnamurthy}},
  \bibnamefont{and} \bibinfo{author}{\bibfnamefont{M.}~\bibnamefont{Barma}},
  \bibinfo{journal}{Phys. Rev. E} \textbf{\bibinfo{volume}{61}},
  \bibinfo{pages}{6337} (\bibinfo{year}{2000}{\natexlab{a}}).

\bibitem[{\citenamefont{Rajesh}(2004)}]{rajesh2004}
\bibinfo{author}{\bibfnamefont{R.}~\bibnamefont{Rajesh}},
  \bibinfo{journal}{Phys. Rev. E} \textbf{\bibinfo{volume}{69}},
  \bibinfo{pages}{036128} (\bibinfo{year}{2004}).

\bibitem[{\citenamefont{Majumdar
  et~al.}(2000{\natexlab{b}})\citenamefont{Majumdar, Krishnamurthy, and
  Barma}}]{MKB2}
\bibinfo{author}{\bibfnamefont{S.~N.} \bibnamefont{Majumdar}},
  \bibinfo{author}{\bibfnamefont{S.}~\bibnamefont{Krishnamurthy}},
  \bibnamefont{and} \bibinfo{author}{\bibfnamefont{M.}~\bibnamefont{Barma}},
  \bibinfo{journal}{J. Stat. Phys.} \textbf{\bibinfo{volume}{99}},
  \bibinfo{pages}{1} (\bibinfo{year}{2000}{\natexlab{b}}).

\bibitem[{\citenamefont{Rajesh and Majumdar}(2001)}]{rajeshmkb1}
\bibinfo{author}{\bibfnamefont{R.}~\bibnamefont{Rajesh}} \bibnamefont{and}
  \bibinfo{author}{\bibfnamefont{S.~N.} \bibnamefont{Majumdar}},
  \bibinfo{journal}{Phys. Rev. E} \textbf{\bibinfo{volume}{63}},
  \bibinfo{pages}{036114} (\bibinfo{year}{2001}).

\bibitem[{\citenamefont{Zaboronski}(2001)}]{zaboronski2001}
\bibinfo{author}{\bibfnamefont{O.}~\bibnamefont{Zaboronski}},
  \bibinfo{journal}{Phys. Lett. A} \textbf{\bibinfo{volume}{281}},
  \bibinfo{pages}{119} (\bibinfo{year}{2001}).

\end{thebibliography}

\end{document}